\pdfoutput=1

\documentclass[11pt]{article}

\usepackage[]{ACL2023}

\usepackage{times}
\usepackage{latexsym}

\usepackage[T1]{fontenc}

\usepackage[utf8]{inputenc}

\usepackage{microtype}

\usepackage{inconsolata}

\newcommand{\ourmodel}{ReContriever\xspace}
\usepackage{makecell}

\usepackage{times}
\usepackage{latexsym}

\usepackage[normalem]{ulem}
\usepackage{mathrsfs}
\usepackage{amsthm}
\usepackage{amsmath}
\usepackage{amssymb, bm}
\usepackage{multirow}
\usepackage{booktabs}
\usepackage{enumitem}
\usepackage{times}
\usepackage{latexsym}
\usepackage{array}
\usepackage{multirow}
\usepackage{graphicx}
\usepackage{amsfonts}
\usepackage{subcaption}
\usepackage{xspace}
\usepackage{tabularx}
\usepackage{algorithm}
\usepackage{arydshln}
\usepackage{algpseudocode}
\usepackage{colortbl}

\definecolor{mypurple}{RGB}{111,61,121}
\definecolor{myblue}{RGB}{46,88,180}
\definecolor{myred}{RGB}{181,68,106}
\definecolor{textorange}{RGB}{237,125,49}
\definecolor{textblue}{RGB}{46,117,181}
\definecolor{textgreen}{RGB}{112,173,71}

\title{Unsupervised Dense Retrieval with \\ Relevance-Aware Contrastive Pre-Training}

\author{Yibin Lei\textsuperscript{1}\thanks{\;\;Work done when Yibin Lei was interning at JD Explore Academy.}\;, Liang Ding\textsuperscript{2}\thanks{\;\;Corresponding author}\;, Yu Cao\textsuperscript{3}, Chantong Zan\textsuperscript{4}, Andrew Yates\textsuperscript{1}, Dacheng Tao\textsuperscript{2}\\
\textsuperscript{1}{University of Amsterdam} \quad
\textsuperscript{2}{JD Explore Academy} \\
\textsuperscript{3}{Tencent IEG} \quad
\textsuperscript{4}{China University of Petroleum (East China)} \\
{\tt \{y.lei, a.c.yates\}@uva.nl, \{liangding.liam, dacheng.tao\}@gmail.com} \\
{\tt rainyucao@tencent.com, zanct@s.upc.edu.cn} 
}

\begin{document}
\maketitle
\begin{abstract}
Dense retrievers have achieved impressive performance, but their demand for abundant training data limits their application scenarios. 
Contrastive pre-training, which constructs pseudo-positive examples from unlabeled data, has shown great potential to solve this problem. 
However, the pseudo-positive examples crafted by data augmentations can be irrelevant. To this end, we propose relevance-aware contrastive learning. It takes the intermediate-trained model itself as an imperfect oracle to estimate the relevance of positive pairs and adaptively weighs the contrastive loss of different pairs according to the estimated relevance. 
Our method consistently improves the SOTA unsupervised Contriever model~\citep{contriever} on the BEIR and open-domain QA retrieval benchmarks. Further exploration shows that our method can not only beat BM25 after further pre-training on the target corpus but also serves as a good few-shot learner. Our code is publicly available at \url{https://github.com/Yibin-Lei/ReContriever}.
\end{abstract}

\section{Introduction}
Dense retrievers, which estimate the relevance between queries and passages in the dense embedding space, have achieved impressive performance in various applications, including web search~\citep{baidu_search} and open-domain question answering~\citep{dpr}. One key factor for the success of dense retrievers is a large amount of human-annotated training data, e.g., MS-MARCO~\citep{msmarco} with above 500,000 examples. However, a recent study~\citep{beir} shows that even trained with enormous labeled data, dense retrievers still suffer from a generalization issue, where they perform relatively poorly on novel domains in comparison to BM25. Meanwhile, collecting human-annotated data for new domains is always hard and expensive. Thus improving dense retrievers with limited annotated data becomes essential, considering the significant domain variations of practical retrieval tasks.

\begin{figure}[t]
\setlength{\abovecaptionskip}{0.1cm}
\setlength{\belowcaptionskip}{-0.1cm}
    \centering
    \includegraphics[width=0.85\linewidth]{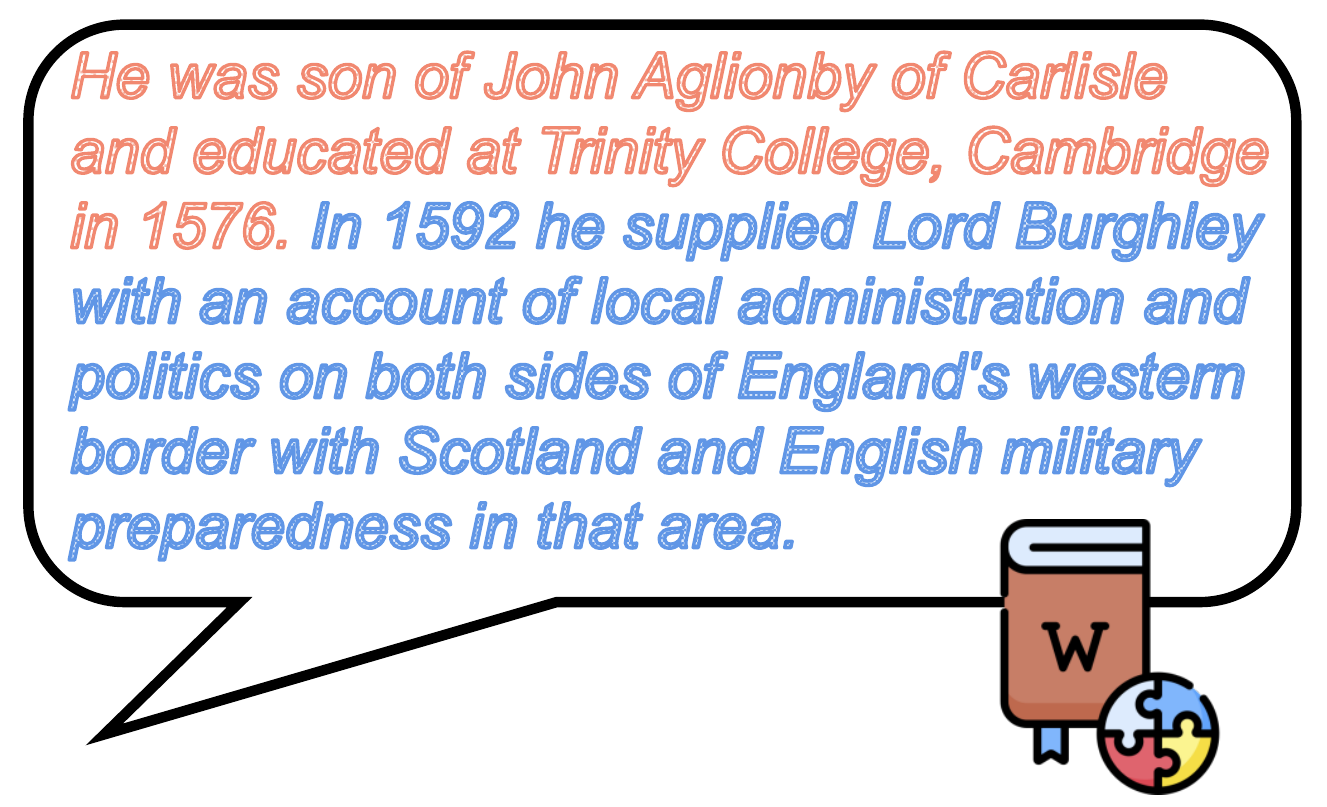}
    \caption{A text snippet from Wikipedia, where two nearby sentences are quite irrelevant. Random cropping may lead to a false positive query-passage pair.}
    \label{fig:illustration}
\end{figure}

Contrastive pre-training, which first generates pseudo-positive examples from a universal corpus and then utilizes them to contrastively pre-train retrievers, has shown impressive performance without any human annotations~\cite{ict, simcse,cocodenser,spider,contriever}. For instance, Contriever~\cite{contriever} crafts relevant query-passage pairs by randomly cropping two random spans within the same document. However, owing to the high information density of texts, even nearby sentences in a document can be very irrelevant, as shown in Figure~\ref{fig:illustration}. These false positive samples may mislead the model to pull unrelated texts together in the embedding space and further harm the validity of representations. 

Motivated by recent findings in computer vision that pre-training performance can be greatly boosted by reducing the effect of such false positives~\cite{betterviews, objectawarecrop}, we propose Relevance-Aware Contrastive Retriever (\ourmodel). 
At each training step, we utilize the trained models at the current step itself to estimate the relevance of all the positives. Then the losses of different positive pairs are adaptively weighed using the estimated relevance, i.e., the pairs that receive higher relevance scores obtain higher weight. Moreover, simply applying lower weights to irrelevant pairs will result in insufficient usage of data, since many documents will contribute less to training. Therefore, we also introduce a one-document-multiple-pair strategy that generates multiple positive pairs from a single document, with a pair-weighting process conducted among samples originating from a single document. Such an operation makes sure that the model can learn positive knowledge from every document in the corpus.

To summarize, our contributions in this paper are three-fold: 
1) We propose relevance-aware contrastive learning for dense retrieval pre-training, which aims to reduce the false positive problem.
2) Experiments show our method brings consistent improvements to the SOTA unsupervised Contriver model on 10/15 tasks on the BEIR benchmark and three representative open-domain QA retrieval datasets.  
3) Further explorations show that our method works well given no or limited labeled data. Specifically, on 4 representative domain-specialized datasets it outperforms BM25 when only unsupervised pre-training on the target corpora, and with only a few annotated samples its accuracy can be on par with DPR~\cite{dpr} which is trained on thousands of annotated examples.

\begin{table*}[t]
\centering
\small
\setlength{\tabcolsep}{2pt}
\setlength{\abovecaptionskip}{0.1cm}
\setlength{\belowcaptionskip}{-0.1cm}
\scalebox{1}{
\begin{tabular}{@{}l  cccccccc} 
  \toprule
           DATASET          & BM25   &  BERT   &   SimCSE & RetroMAE     &   coCondenser     &  Contriever  & \makecell[c]{Contriever \\ (reproduced)}    & \ourmodel   \\
    \midrule
    MS MARCO    &  \textbf{22.8}   & 0.6    &     8.8       &   4.5     &     7.7       &     20.6                  &     21.1
                &       ~~$21.8^\dag$ \\
    Trec-COVID  &   \textbf{65.6}   & 16.6   &   38.6     &    20.4     &    17.3        &    27.4                    &    42.0
                &     40.5   \\
    NFCorpus    &   \textbf{32.5}  &  2.5    &   14.0      &   15.3      &     14.4       &    31.7                   &     30.0
                 &      ~~$31.9^\dag$  \\
    NQ          &   \textbf{32.9}  &  2.7    &   12.6    &   3.4      &     3.9         &    25.4                    &     29.5 
                &        ~~$31.0^\dag$ \\  
    HotpotQA    &   \textbf{60.3}  &  4.9    &   23.3    &   25.0     &    24.4           &    48.1                    &     44.1  
                &      ~~$50.1^\dag$ \\ 
    FiQA-2018   &   23.6  &  1.4    &   14.8    &  9.3     &    5.2            &    24.5                     &      \textbf{26.2}  
                &      \textbf{26.2} \\ 
    ArguAna     &   31.5  &  23.1    &    \textbf{45.6}    &   37.6     &     34.5       &    37.9                  &     43.4  
                &        39.8 \\ 
    Touche-2020 &   \textbf{36.7}  &  3.4    &    11.6    &  1.9     &    3.0          &    16.7                  &       16.7
                &        16.6 \\ 
    CQADupStack &   \textbf{29.9}  &  2.5    &    20.2    &    17.0    &     9.8          &    28.4                  &     28.4  
                &        ~~$28.7^\dag$ \\ 
    Quora       &   78.9  &  3.9    &    81.5    &  69.0    &     66.7         &    83.5                    &     83.6  
                &        ~~$\textbf{84.3}^\dag$ \\ 
    DBPedia     &   \textbf{31.3}  &  3.9    &    13.7    &     4.6  &     15.1           &    29.2                    &     27.6  
                &   ~~$29.3^\dag$ \\ 
    SCIDOCS     &   \textbf{15.8}  &  2.7    &    7.4    &   7.4      &     1.9          &    14.9                    &      15.0  
                &      ~~$15.6^\dag$ \\ 
    FEVER       &   \textbf{75.3}  &  4.9    &    20.1    &  7.1     &    25.3          &    68.2                    &     66.9
                &        ~~$68.9^\dag$ \\ 
  Climate-fever &   \textbf{21.3}  &  4.1    &    17.6    &  4.4    &    9.8         &    15.5                    &     15.6  
                &        15.6 \\ 
    SciFact     &   \textbf{66.5}  &  9.8    &    38.5    &   53.1    &    48.1          &    64.9                    &     65  
                &        66.4 \\ 
    \midrule
    \textbf{Avg}     &   \textbf{41.7}  &  8.7    &    24.6    &   18.7     &     8.9           &       35.8                    &     37.0  &    37.8 \\    
    \textbf{Avg Rank} & \textbf{1.9}   & 7.9    &   4.9    &  6.1    &   6.3          &     3.4                  &  2.7    &  2.2   \\
\bottomrule
\end{tabular}}
\caption{\textbf{NDCG@10 of BEIR Benchmark}. All models are \textbf{unsupervised trained without any human-annotated data}. \textbf{Bold} indicates the best result. The average and rank across the entire benchmark are included. Four datasets are excluded  because of their licenses. ``$^{\dagger}$'' means ReContriever performs significantly better than our reproduced Contriever, as determined by a t-test with p-value 0.05 as threshold.}
\label{tab:unsup_beir}
\end{table*}

\section{Method}
\subsection{Preliminary}
In this section, we briefly describe the bi-encoder structure used in dense retrieval and the SOTA Contriever model, on which we build our model.

\paragraph{Bi-Encoder Structure}
Dense retrievers are always a bi-encoder composed of two separate encoders to transform the query and document into a single vector each. The relevance score is obtained by computing the similarity (e.g., inner-product) between the encoded vectors of queries and documents. The typical way to train a dense retriever is using a contrastive loss that aims to pull relevant passages closer to the query and irrelevant passages farther in the embedding space. For each query, the training data involves one positive passage labeled by annotators and a pool of negative passages, which are usually random passages in the corpus.

\paragraph{Contriever}
It crafts pseudo-positive pairs by randomly cropping two spans of the same document. As negative texts have shown to be a key to the success of retrieval training~\citep{ance}, Contriever also applies the MoCo mechanism~\cite{moco} to utilize negatives in the previous batches to increase the number of negatives. 
These two factors make Contriever obtain significant decent performance without any human annotations.

\subsection{Relevance-Aware Contrastive Learning}
We start by 1) producing a larger number of positives (\textbf{\textit{one-document-multi-pair}}) and 2) forcing the model to pay more attention to the ones with higher relevance (\textbf{\textit{relevance-aware contrastive loss}}).

\paragraph{One-Document-Multi-Pair}
Given a text snippet $T$, previous pre-training methods always craft only one positive (query-passage) pair $(q,d^+)$.
To exploit $T$ more effectively, our one-document-multi-pair strategy generates $n$ positive pairs, denoting as $\{(q, d^+_1), (q, d^+_2), \ldots, (q, d^+_n)\}$, from $T$ by repeating the procedure several times. 
We keep the query $q$ unchanged to ensure the relevance comparison is fair among pairs within the same snippet, which is used in our following step. Building upon Contirever, we craft $n$ pairs by random cropping $n+1$ spans and setting 1 span as the fixed query for the left $n$ spans. And it is easy to extend this strategy to other contrastive pre-training methods. 

\paragraph{Relevance-Aware Contrastive Loss}
The ordinary contrastive loss for training dense retrievers is the InfoNCE loss. Given a positive pair $(q,d^+)$ and a negative pool $\{d^-_i\}_{i=1 . . D}$,    $\text{InfoNCE} \left(q, d^{+}\right)$ is computed by:
\begin{small}
\begin{equation}
     -\ \text{log} \ 
     \frac{\exp \left(s\left(q, d^{+}\right) / \tau\right)}{\exp \left(s\left(q, d^{+}\right) / \tau\right)+\sum_{i=1}^D \exp \left(s\left(q, d^-_i\right) / \tau\right)},
\end{equation}
\end{small}
where $s(\cdot)$ and $\tau$ denote the similarity function and temperature parameter. Then the overall loss of a batch is usually the average across all the $m\times n$ positive pairs from $m$ snippets: $L= {\frac{1}{m n}}\sum_{i=1}^m \sum_{j=1}^n \text{InfoNCE}(q_i, d^+_{ij})$.

The relevance-aware contrastive loss aims to force the model to focus more on true positive pairs by 1) utilizing trained model $\theta$ at present itself as an imperfect oracle
to compute the relevance score $s_\theta(q,d^+)$ between all pairs;
and 2) adaptively assigning weights to different pairs according to the estimated relevance. Then the relevance-aware contrastive loss $L_\text{relevance}$ can be expressed as:\footnote{Equation (\ref{relevance_loss}) will be invalid when $s_\theta\left(q_i, d_{i j}^{+}\right)$ is negative. In the preliminary study, we found the value is always positive and thus ignore this special case for simplicity.}
\begin{equation}
\begin{small}
\label{relevance_loss}
{\frac{1}{m}}\sum_{i=1}^m \sum_{j=1}^n \frac{s_\theta\left(q_i, d_{i j}^{+}\right)}{\sum_{k=1}^n s_\theta\left(q_i, d_{i k}^{+}\right)} 
\text{InfoNCE}(q_i, d^+_{ij}).
\end{small}
\end{equation}
In this way, for each text snippet, positive pairs with more confidence to be relevant will thus be more focused on by the model, or vice versa.

\section{Experiments}
In this section, we evaluate our model in several settings after describing our experimental setup. We consider unsupervised retrieval performance and two practical use cases: further pre-training on the target domain and few-short retrieval. We then conduct an ablation study to separate the impact of our method's two components.

\subsection{Setup}\label{sec:setup}
\begin{itemize}[wide=0\parindent, noitemsep, topsep=0pt]
    \item \textbf{Datasets} We evaluate retrieval models on the BEIR~\cite{beir} benchmark and three representative open-domain QA retrieval benchmarks: Natural Questions (NQ;~\cite{nq}), TriviaQA~\cite{triviaqa} and WebQuestions~\cite{webquestions}.     
    \item \textbf{Baselines} We compare our model with two types of unsupervised models, namely models based on contrastive pre-training and on auto-encoding pre-training. The former models include SimCSE~\cite{simcse}, coCondenser~\cite{cocodenser}, Spider~\cite{spider} and Contriever~\cite{contriever}. The latter category includes the recently proposed RetroMAE~\cite{retromae}. BM25~\cite{bm25} and uncased BERT-base model~\cite{bert} are also involved for reference. We use the official checkpoints for evaluation.
    \item \textbf{Implementation Details} We apply our method to the SOTA Contriever model and use its default settings. The pre-training data is a combination of Wikipedia and CCNet~\citep{ccnet}, same as Contriever. We generate 4 positive pairs for each document. Refer to Appendix~\ref{app:imlementations} for more details.
    
    We conduct a t-test with p-value 0.05 as threshold to compare the performance of ReContriever and our reproduced Contriever.
\end{itemize}

\begin{table*}[ht]
\setlength{\abovecaptionskip}{0.1cm}
\setlength{\tabcolsep}{4pt}
\centering
\small
\begin{tabular}{lccccccccc}
\toprule
\multirow{2}{15pt}{Model} & 
\multicolumn{3}{c}{NQ} & \multicolumn{3}{c}{TriviaQA} & \multicolumn{3}{c}{WQ}
\\ \cmidrule(lr){2-4} \cmidrule(lr){5-7} \cmidrule(lr){8-10} 
& Top-5 & Top-20 & Top-100 & Top-5 & Top-20 & Top-100& Top-5 & Top-20 & Top-100  \\
\midrule
 & \multicolumn{9}{c}{\textit{Supervised Model}}  \\
 \midrule
DPR   & - & 78.4 & 85.4 & - & 79.4 & 85.0 & - & 73.2 & 81.4 \\ 

\midrule
& \multicolumn{9}{c}{\textit{Unsupervised Models}}  \\
\midrule
BM25 & 43.8 & 62.9 & 78.3 & \textbf{66.3} & \textbf{76.4} & 83.2 & 41.8 & 62.4 & 75.5 \\ 
RetroMAE & 23.0 & 40.1 & 58.8 & 47.0 & 61.4 & 74.2 & 25.8 & 43.8 & 62.3 \\
SimCSE  & 5.4 & 11.5 & 23.0 & 3.7 & 7.6 & 17.0 & 3.3 & 8.7 & 19.4 \\
coCondenser  & 28.9 & 46.8 & 63.5 & ~~7.5 & 13.8 & 24.3 & 30.2 & 50.7 & 68.7 \\
Spider  & 49.6 & 68.3 & 81.2 & 63.6 & 75.8 & 83.5 & 46.8 & 65.9 & 79.7 \\ 
Contriever  & 47.3 & 67.8 &  80.6 & 59.5 &  73.9 &  82.9 & 43.5 & 65.7 & 80.1 \\
Contriever (reproduced) & 48.9 & 68.3 & 81.4 & 61.2 & 74.6 & 83.4 & 47.0 & 67.0 & 80.5 \\
\ourmodel  & ~~\textbf{50.3}$^\dag$ & ~~\textbf{69.4}$^\dag$ & ~~\textbf{82.6}$^\dag$ & ~~63.4$^\dag$ & ~~75.9$^\dag$ & ~~\textbf{84.1}$^\dag$ & \textbf{48.3} & \textbf{68.0} &  \textbf{81.1}\\ 
\bottomrule
\end{tabular}
\caption{\textbf{Recall of open-domain retrieval benchmarks}. \textbf{Bold}: the best results across unsupervised models. ``$^{\dagger}$'' means ReContriever performs significantly better than our reproduced Contriever, as determined by a t-test with p-value 0.05 as threshold.}
\label{tab:open_domain}
\end{table*}

\subsection{Main Results}
\label{unsup_retrieval}
\subsubsection{BEIR}
The NDCG@10 of \ourmodel and other fully unsupervised models across 15 public datasets of BEIR are shown in Table~\ref{tab:unsup_beir}.  
\ourmodel achieves consistent improvements over Contriever on 10/15 datasets, with a significant improvement observed in 9 of those datasets. Notably, it also only sees very slight decreases on datasets without promotion ({e.g., FiQA, Touche and Climate-Fever with at most -0.1 decrease). Moreover, our method obtains an average rank of 2.2, proving our method to be the best unsupervised dense retriever. BM25 is still a strong baseline under the fully unsupervised scenario, but \ourmodel greatly narrows the gap between dense retrievers and it.

\subsubsection{Open-Domain QA Retrieval}
Table~\ref{tab:open_domain} shows the Recall performance of \ourmodel on open-domain QA retrieval benchmarks, where supervised DPR~\cite{dpr} is involved for reference. Obviously, \ourmodel outperforms BM25 by a large margin except for Recall@5 and Recall@10 on TriviaQA with relatively smaller differences, verifying the effect of our method. Moreover, among all unsupervised methods, \ourmodel obtains the best performance in nearly all cases, especially substantial improvement over Contriever. Our \ourmodel promisingly narrows the gaps between supervised and unsupervised models, making it more valuable.

\begin{table}[ht]
\centering
\footnotesize
\setlength{\tabcolsep}{3.5pt}
\setlength{\abovecaptionskip}{0.1cm}
\setlength{\belowcaptionskip}{-0.1cm}
\begin{tabular}{@{} l ccccc @{}}
\toprule
Model & SciF & SCID & Arg & CQA & Avg. \\
\midrule
BM25 & 66.5 & 15.8 & 31.5 & 29.9 & 35.9 \\
\midrule 
Contriever & 64.9 & 14.9 & 43.4 & 28.4 & 37.9 \\
\quad + corpus pretrain & 66.3 & \textbf{17.1} & 52.4 & 30.6 & 41.6$^{\Uparrow+3.7}$\\
\midrule 
\ourmodel & 66.4  & 15.6 & 39.8 & 28.4 & 37.6\\
\quad + corpus pretrain & \textbf{67.1} & 16.6 & ~~\textbf{54.6}$^{\dagger}$ & \textbf{30.7} & \textbf{42.3}$^{\Uparrow\mathbf{+4.7}}$ \\
\bottomrule
\end{tabular}
\caption{\textbf{NDCG@10 after further pre-training on the target domain corpus}. 
``$^{\Uparrow}$'' denotes the gains of further pre-training.
``$^{\dagger}$'' means ReContriever performs significantly better than our reproduced Contriever.
}
\label{tab:target_pretrian}
\end{table}

\subsection{Practical Use Cases}
\label{sec:practical}
In this section, we explore the applicability of \ourmodel in more practical scenarios\footnote{We report results of our reproduced Contriever as they are slightly better than the original ones~\citep{contriever}.}, where only texts in the target corpus (pre-training on the target domain) or very limited annotated training data (few-shot retrieval) are available.

\begin{table}[ht]
\setlength{\abovecaptionskip}{0.1cm}
\setlength{\belowcaptionskip}{-0.1cm}
\centering
\footnotesize
\begin{tabular}{lccc}
\toprule
\multirow{3}{10pt}{Model} & 
\multicolumn{3}{c}{NQ} 
\\ \cmidrule(lr){2-4} 
& Top-5 & Top-20 & Top-100 \\
\midrule
& \multicolumn{3}{c}{\textit{Reference}}  \\
\midrule
DPR   &  - & 78.4 & 85.4  \\ 
BM25 & 43.8 & 62.9 & 78.3 \\ 
\midrule
& \multicolumn{3}{c}{\textit{8 examples}}  \\
\midrule
Spider & 49.7 & 68.3 & 81.4\\
Contriever & 51.7 & 70.6 & 83.1  \\
\ourmodel  & 52.9 & 71.6 & ~~84.2$^{\dagger}$\\ 
\midrule
& \multicolumn{3}{c}{\textit{32 examples}}  \\
\midrule
Spider & 50.2 & 69.4 & 81.7\\
Contriever & 52.6 & 70.9 & 83.1  \\
\ourmodel  & 53.5 & ~~71.9$^{\dagger}$ & ~~84.7$^{\dagger}$\\ 
\midrule
& \multicolumn{3}{c}{\textit{128 examples}}  \\
\midrule
Spider & 57.0 & 74.3 & 85.3\\
Contriever & 55.1 & 72.4 & 83.7  \\
\ourmodel  & 55.9 & ~~74.1$^{\dagger}$ & ~~85.1$^{\dagger}$\\ 

\bottomrule
\end{tabular}
\caption{\textbf{Few-shot Retrieval on NQ}. Results are report with Recall. ``$^{\dagger}$'' means ReContriever performs significantly better than our reproduced Contriever.}
\label{tab:few-shot}
\end{table}

\begin{table*}[ht]
\centering
\small
\setlength{\tabcolsep}{5pt}
\setlength{\abovecaptionskip}{0.1cm}
\setlength{\belowcaptionskip}{-0.1cm}
\scalebox{0.97}{
\begin{tabular}{@{}l  cccccccccccccccc} 
  \toprule
      Model           & MS MARCO    &   NFCoprus    &   NQ   &   Hotpot    &  FiQA   & Touche  & Quora &  SCIDOCS & Avg.\\
\midrule
Contriever & 19.1 & 25.1 & 26.7 & 43.2 & 23.2 & 18.6 & 82.3 & 14.6 & 31.6 \\
\quad + relevance-aware loss & 0.2 & 2.5 & 0.0 & 0.2 & 0.5 & 0.6 & 57.6 & 14.5 & 9.5\\
\quad + one-document-multiple-pair & 19.9 & 29.5 & 27.5 & 44.2 & 21.9 & 15.9 & 82.8 & 14.5 & 32.0 \\
\midrule
ReContriever  &  20.8 & 28.1 & 29.6 & 49.9 & 23.4 & 18.2 & 83.3 &  14.7 & 33.5\\

\bottomrule

\end{tabular}}
\caption{\textbf{Ablation Study.} Results are reported with NDCG@10.
}
\label{tab:ablation_study}
\end{table*}

\paragraph{Pre-Training on the Target Domain}
Four domain-specialized datasets (SciFact~\citep{scifact} (SciF; citation-prediction), SCIDOCS~\cite{scidocs} (SCID; fact checking), ArguAna~\cite{arguana} (Arg; argument retrieval) and CQADupStack~\cite{cqadupstack} (CQA; StackExchange retrieval) with only corpus available are picked as a testbed, shown in Table~\ref{tab:target_pretrian}. Before further pre-training on the corresponding corpus, \ourmodel underperforms BM25 on 3 of 4 datasets. Surprisingly, after pre-training, \ourmodel is able to consistently beat BM25. Moreover, our model obtains an average +4.7 improvement after further pre-training, which is substantially better than Contriever (+3.7).

\paragraph{Few-Shot Retrieval.}
\label{app:few-shot}
Results on training with limited annotated data of NQ are shown in Table~\ref{tab:few-shot}. Following the same setting, our model trained on 128 samples can perform on par on Recall@100 with DPR which has seen thousands of annotated samples. In addition, when training data is scarce (below 100 examples), \ourmodel still shows stronger few-shot performance compared to Spider and Contriever.

\subsection{Ablation Study}
\label{sec:ablation}
We conduct an ablation study to investigate the contributions of our proposed loss and pairing strategies within \ourmodel, using 100,000 training steps. Solely adding relevance-aware loss means estimating the relevance of N pairs from N documents and then normalizing the relevance among the N pairs within a batch, which slightly differs from equation (\ref{relevance_loss}) that normalizes over 4 pairs from the same document. As shown in Table~\ref{tab:ablation_study}, solely adding relevance-aware contrastive loss to Contriever will lead to a noticeable degeneration, owing to the missing information from the documents with low adjusted weights and the unstable relevance comparison without a fixed query. Applying the one-document-multi-pair strategy can obtain a slight improvement which can be attributed to the effective usage of the unlabeled data. Combining both strategies (i.e., ReContriever) can lead to an obvious improvement, which demonstrates the necessity of both components in our method.

\section{Conclusion}
In this work, we propose \ourmodel to further explore the potential of contrastive pre-training to reduce the demand of human-annotated data for dense retrievers.
Benefiting from multiple positives from the same document as well as relevance-aware contrastive loss, our model achieves remarkable performance under zero-shot cases. Additional results on low data resources further verify its value under various practical scenarios.

\section*{Limitations}
Although \ourmodel narrows the gap between BM25 and unsupervised dense retrievers, it still lags behind BM25 when acting as a general-purpose retriever. This issue may make \ourmodel not directly usable when facing a new domain, thus limiting its practicality. Also, 
as \ourmodel is initialized from the language model BERT$_\text{base}$, there may exist social biases~\cite{bias} in \ourmodel and thus have the risk of offending people from under-represented groups.

\section*{Ethics Statement}
We strictly adhere to the ACL Ethics Policy. This paper focuses on reducing the false positives problem of unsupervised dense retrieval. The datasets used in this paper are publicly available and have been widely adopted by researchers. We ensure that the findings and conclusions of this paper are reported accurately and objectively.

\bibliography{anthology,custom}

\begin{thebibliography}{25}
\expandafter\ifx\csname natexlab\endcsname\relax\def\natexlab#1{#1}\fi

\bibitem[{Bajaj et~al.(2016)Bajaj, Campos, Craswell, Deng, Gao
  et~al.}]{msmarco}
Payal Bajaj, Daniel Campos, Nick Craswell, Li~Deng, Jianfeng Gao, et~al. 2016.
\newblock \href {https://arxiv.org/abs/1611.09268} {Ms marco: A human generated
  machine reading comprehension dataset}.
\newblock \emph{arXiv preprint}.

\bibitem[{Berant et~al.(2013)Berant, Chou, Frostig, and Liang}]{webquestions}
Jonathan Berant, Andrew~K. Chou, Roy Frostig, and Percy Liang. 2013.
\newblock Semantic parsing on freebase from question-answer pairs.
\newblock In \emph{EMNLP}.

\bibitem[{Cohan et~al.(2020)Cohan, Feldman, Beltagy, Downey, and
  Weld}]{scidocs}
Arman Cohan, Sergey Feldman, Iz~Beltagy, Doug Downey, and Daniel Weld. 2020.
\newblock \href {https://aclanthology.org/2020.acl-main.207} {{SPECTER}:
  Document-level representation learning using citation-informed transformers}.
\newblock In \emph{ACL}.

\bibitem[{Devlin et~al.(2019)Devlin, Chang, Lee, and Toutanova}]{bert}
Jacob Devlin, Ming-Wei Chang, Kenton Lee, and Kristina Toutanova. 2019.
\newblock \href {https://aclanthology.org/N19-1423} {{BERT}: Pre-training of
  deep bidirectional transformers for language understanding}.
\newblock In \emph{NAACL}.

\bibitem[{Gao and Callan(2022)}]{cocodenser}
Luyu Gao and Jamie Callan. 2022.
\newblock \href {https://aclanthology.org/2022.acl-long.203} {Unsupervised
  corpus aware language model pre-training for dense passage retrieval}.
\newblock In \emph{ACL}.

\bibitem[{Gao et~al.(2021)Gao, Yao, and Chen}]{simcse}
Tianyu Gao, Xingcheng Yao, and Danqi Chen. 2021.
\newblock \href {https://aclanthology.org/2021.emnlp-main.552/} {{SimCSE}:
  Simple contrastive learning of sentence embeddings}.
\newblock In \emph{EMNLP}.

\bibitem[{He et~al.(2020)He, Fan, Wu, Xie, and Girshick}]{moco}
Kaiming He, Haoqi Fan, Yuxin Wu, Saining Xie, and Ross Girshick. 2020.
\newblock Momentum contrast for unsupervised visual representation learning.
\newblock In \emph{CVPR}.

\bibitem[{Hoogeveen et~al.(2015)Hoogeveen, Verspoor, and Baldwin}]{cqadupstack}
Doris Hoogeveen, Karin~M. Verspoor, and Timothy Baldwin. 2015.
\newblock \href {http://doi.acm.org/10.1145/2838931.2838934} {Cqadupstack: A
  benchmark data set for community question-answering research}.
\newblock In \emph{ADCS}.

\bibitem[{Izacard et~al.(2022)Izacard, Caron, Hosseini, Riedel, Bojanowski,
  Joulin, and Grave}]{contriever}
Gautier Izacard, Mathilde Caron, Lucas Hosseini, Sebastian Riedel, Piotr
  Bojanowski, Armand Joulin, and Edouard Grave. 2022.
\newblock \href {https://openreview.net/forum?id=jKN1pXi7b0} {Unsupervised
  dense information retrieval with contrastive learning}.
\newblock \emph{TMLR}.

\bibitem[{Joshi et~al.(2017)Joshi, Choi, Weld, and Zettlemoyer}]{triviaqa}
Mandar Joshi, Eunsol Choi, Daniel Weld, and Luke Zettlemoyer. 2017.
\newblock \href {https://aclanthology.org/P17-1147} {{T}rivia{QA}: A large
  scale distantly supervised challenge dataset for reading comprehension}.
\newblock In \emph{ACL}.

\bibitem[{Karpukhin et~al.(2020)Karpukhin, Oguz, Min, Lewis, Wu, Edunov, Chen,
  and Yih}]{dpr}
Vladimir Karpukhin, Barlas Oguz, Sewon Min, Patrick Lewis, Ledell Wu, Sergey
  Edunov, Danqi Chen, and Wen-tau Yih. 2020.
\newblock \href {https://aclanthology.org/2020.emnlp-main.550} {Dense passage
  retrieval for open-domain question answering}.
\newblock In \emph{EMNLP}.

\bibitem[{Kwiatkowski et~al.(2019)Kwiatkowski, Palomaki, Redfield, Collins,
  Parikh, Alberti, Epstein, Polosukhin, Devlin, Lee, Toutanova, Jones, Kelcey,
  Chang, Dai, Uszkoreit, Le, and Petrov}]{nq}
Tom Kwiatkowski, Jennimaria Palomaki, Olivia Redfield, Michael Collins, Ankur
  Parikh, Chris Alberti, Danielle Epstein, Illia Polosukhin, Jacob Devlin,
  Kenton Lee, Kristina Toutanova, Llion Jones, Matthew Kelcey, Ming-Wei Chang,
  Andrew~M. Dai, Jakob Uszkoreit, Quoc Le, and Slav Petrov. 2019.
\newblock \href {https://aclanthology.org/Q19-1026} {Natural questions: A
  benchmark for question answering research}.
\newblock \emph{TACL}.

\bibitem[{Lee et~al.(2019)Lee, Chang, and Toutanova}]{ict}
Kenton Lee, Ming-Wei Chang, and Kristina Toutanova. 2019.
\newblock \href {https://aclanthology.org/P19-1612} {Latent retrieval for
  weakly supervised open domain question answering}.
\newblock In \emph{ACL}.

\bibitem[{Liu et~al.(2021)Liu, Lu, Cheng, Shi, Wang, Cheng, and
  Yin}]{baidu_search}
Yiding Liu, Weixue Lu, Suqi Cheng, Daiting Shi, Shuaiqiang Wang, Zhicong Cheng,
  and Dawei Yin. 2021.
\newblock \href {https://doi.org/10.1145/3447548.3467149} {Pre-trained language
  model for web-scale retrieval in baidu search}.
\newblock In \emph{KDD}.

\bibitem[{Mishra et~al.(2022)Mishra, Shah, Bansal, Anjaria, Jagannatha, Sharma,
  Jacobs, and Krishnan}]{objectawarecrop}
Shlok~Kumar Mishra, Anshul Shah, Ankan Bansal, Janit~K Anjaria,
  Abhyuday~Narayan Jagannatha, Abhishek Sharma, David Jacobs, and Dilip
  Krishnan. 2022.
\newblock \href {https://openreview.net/forum?id=WXgJN7A69g} {Object-aware
  cropping for self-supervised learning}.
\newblock \emph{TMLR}.

\bibitem[{Peng et~al.(2022)Peng, Wang, Zhu, Wang, and You}]{betterviews}
Xiangyu Peng, Kai Wang, Zheng Zhu, Mang Wang, and Yang You. 2022.
\newblock Crafting better contrastive views for siamese representation
  learning.
\newblock In \emph{CVPR}.

\bibitem[{Ram et~al.(2022)Ram, Shachaf, Levy, Berant, and Globerson}]{spider}
Ori Ram, Gal Shachaf, Omer Levy, Jonathan Berant, and Amir Globerson. 2022.
\newblock \href {https://aclanthology.org/2022.naacl-main.193} {Learning to
  retrieve passages without supervision}.
\newblock In \emph{NAACL}.

\bibitem[{Robertson and Zaragoza(2009)}]{bm25}
Stephen Robertson and Hugo Zaragoza. 2009.
\newblock \href {https://doi.org/10.1561/1500000019} {The probabilistic
  relevance framework: Bm25 and beyond}.
\newblock \emph{Found. Trends Inf. Retr.}

\bibitem[{Thakur et~al.(2021)Thakur, Reimers, R{\"u}ckl{\'e}, Srivastava, and
  Gurevych}]{beir}
Nandan Thakur, Nils Reimers, Andreas R{\"u}ckl{\'e}, Abhishek Srivastava, and
  Iryna Gurevych. 2021.
\newblock \href {https://openreview.net/forum?id=wCu6T5xFjeJ} {{BEIR}: A
  heterogeneous benchmark for zero-shot evaluation of information retrieval
  models}.
\newblock In \emph{NeurIPS Datasets and Benchmarks}.

\bibitem[{Wachsmuth et~al.(2018)Wachsmuth, Syed, and Stein}]{arguana}
Henning Wachsmuth, Shahbaz Syed, and Benno Stein. 2018.
\newblock \href {https://aclanthology.org/P18-1023} {Retrieval of the best
  counterargument without prior topic knowledge}.
\newblock In \emph{ACL}.

\bibitem[{Wadden et~al.(2020)Wadden, Lin, Lo, Wang, van Zuylen, Cohan, and
  Hajishirzi}]{scifact}
David Wadden, Shanchuan Lin, Kyle Lo, Lucy~Lu Wang, Madeleine van Zuylen, Arman
  Cohan, and Hannaneh Hajishirzi. 2020.
\newblock \href {https://aclanthology.org/2020.emnlp-main.609/} {Fact or
  fiction: Verifying scientific claims}.
\newblock In \emph{EMNLP}.

\bibitem[{Wenzek et~al.(2020)Wenzek, Lachaux, Conneau et~al.}]{ccnet}
Guillaume Wenzek, Marie-Anne Lachaux, Alexis Conneau, et~al. 2020.
\newblock \href {https://aclanthology.org/2020.lrec-1.494} {{CCN}et: Extracting
  high quality monolingual datasets from web crawl data}.
\newblock In \emph{LREC}.

\bibitem[{Xiao et~al.(2022)Xiao, Liu, Yingxia, and Zhao}]{retromae}
Shitao Xiao, Zheng Liu, Shao Yingxia, and Cao Zhao. 2022.
\newblock \href {https://arxiv.org/abs/2205.12035} {Retromae: Pre-training
  retrieval-oriented language models via masked auto-encoder}.
\newblock In \emph{EMNLP}.

\bibitem[{Xiong et~al.(2021)Xiong, Xiong, Li, Tang, Liu, Bennett, Ahmed, and
  Overwijk}]{ance}
Lee Xiong, Chenyan Xiong, Ye~Li, Kwok-Fung Tang, Jialin Liu, Paul~N. Bennett,
  Junaid Ahmed, and Arnold Overwijk. 2021.
\newblock \href {https://openreview.net/forum?id=zeFrfgyZln} {Approximate
  nearest neighbor negative contrastive learning for dense text retrieval}.
\newblock In \emph{ICLR}.

\bibitem[{Zhao et~al.(2017)Zhao, Wang, Yatskar, Ordonez, and Chang}]{bias}
Jieyu Zhao, Tianlu Wang, Mark Yatskar, Vicente Ordonez, and Kai-Wei Chang.
  2017.
\newblock \href {https://aclanthology.org/D17-1323} {Men also like shopping:
  Reducing gender bias amplification using corpus-level constraints}.
\newblock In \emph{EMNLP}.

\end{thebibliography}
\bibliographystyle{acl_natbib}

\appendix

\section{Implementation Details}
\label{app:imlementations}
\paragraph{Basic Infrastructure}
Basic backbones of our implementation involve Pytorch\footnote{\url{https://pytorch.org/}} and Huggingface-Transformers\footnote{\url{https://github.com/huggingface/transformers}}. We build our code upon the released code of Contriever\footnote{\url{https://github.com/facebookresearch/contriever}}. Models are evaluated using evaluation scripts provided by the BEIR\footnote{\url{https://github.com/beir-cellar/beir}} (for BEIR evaluation) and Spider\footnote{\url{https://github.com/oriram/spider}} (for open-domain QA retrieval evaluation) GitHub repositories. The pre-training experiments are conducted on 16 NVIDIA A100 GPUs and the few-shot experiments are conducted on a single NVIDIA A100 GPU. We report the results on a single run with a fixed random seed 0 (same as the setting of Contriever).

\paragraph{Details of \ourmodel}
    Following the default settings of Contriever, we pre-train \ourmodel for 500,000 steps with a batch size of 2048, initializing from the uncased BERT$_\text{base}$ model with 110 million parameters. The pre-training data is a combination of Wikipedia and CCNet~\citep{ccnet}. The learning rate is set to $5 \cdot 10^{-5}$ with a warm-up for the first 20,000 steps and a linear decay for the remaining steps. Average pooling over the whole sequence is used for obtaining the final representation of the query or document. For each document, we generate 4 positive pairs. 
    
    For experiments on target domain pre-training, we initialize the model from our pre-trained Contriever/ReContriever checkpoints. To avoid overfitting, the models are further pre-trained with 5000 warm-up steps to a learning rate of $1.25 \cdot 10^{-7}$ on all 4 picked datasets with a batch size of 1024 on 8 NVIDIA A100 GPUs.

    For few-shot retrieval experiments, we adopt the training procedure from~\cite{dpr}: exploiting BM25 negatives and not including negatives mined by the model itself~\cite{ance} for few-shot fine-tuning. The hyper-parameters are directly borrowed from~\cite{dpr} except for the batch size and number of training epochs}. We fine-tune all the models with 80 epochs. For 8 examples, the batch size is set to 8. The batch size is 32 when there are 32 or 128 examples.

\section{Dataset Statistics}
Details about the number of examples in the there open-domain QA retrieval datasets are shown in Table~\ref{tab:data_stat}.

\begin{table}[hbt]
\setlength{\abovecaptionskip}{0.1cm}
\setlength{\belowcaptionskip}{-0.2cm}
\centering
\small
\setlength{\tabcolsep}{3pt}
\begin{tabular}{@{} l ccc @{}}
\toprule
Dataset & Train & Dev & Test  \\
\midrule
NQ & 58880 & 8757 & 3610 \\
TriviaQA & 60413 & 8837 & 11313\\
WQ & 2474 & 361 & 2032\\
\bottomrule
\end{tabular}
\caption{\textbf{Statistics of Open-Domain QA Retrieval Datasets} 
}
\label{tab:data_stat}
\end{table}

\end{document}